\documentclass{article}
\usepackage{spconf,amsmath,graphicx,hyperref}
\usepackage{subfiles}
\usepackage{siunitx}
\usepackage{bm}

\title{LOW-POWER END-TO-END COCHLEAR IMPLANT SPEECH DENOISING\\WITH SPIKING NEURAL NETWORKS}
\name{Ludovic Boulanger, Sean U N Wood\thanks{This work was enabled in part by support from the Quebec Research Fund (https://doi.org/10.69777/365045), Ministry of Economy, Innovation and Energy, Calcul Québec (calculquebec.ca), the Digital Research Alliance of Canada (alliancecan.ca), and the University of Sherbrooke.}\thanks{© 2026 IEEE. Personal use of this material is permitted. Permission from IEEE must be obtained for all other uses, in any current or future media, including reprinting/republishing this material for advertising or promotional purposes, creating new collective works, for resale or redistribution to servers or lists, or reuse of any copyrighted component of this work in other works. DOI: https://doi.org/10.1109/ICASSP55912.2026.11464775}}
\address{Department of Electrical and Computer Engineering\\University of Sherbrooke, QC, Canada}

\let\oldbibliography\thebibliography
\renewcommand{\thebibliography}[1]{%
  \oldbibliography{#1}%
  \ninept 
  \setlength{\itemsep}{0pt}
}
\begin{document}
\topmargin=0mm
%
\maketitle
\begin{abstract}
Cochlear implants (CI) restore hearing for individuals with severe to profound hearing loss. However, CI users often struggle to understand speech in noisy environments. Deep neural networks (DNN) have shown promise in enhancing speech for CI users, yet their high energy demands make them non-ideal for low-power CI processors. Spiking neural networks (SNN), on the other hand, offer comparable performance with significantly lower energy consumption. Hence, we propose a novel SNN inspired by the Deep ACE architecture that simultaneously performs speech enhancement and CI coding. Our model achieves competitive vocoded short-time objective intelligibility (VSTOI) and signal-to-noise ratio improvement (SNRi) scores compared to Deep ACE, while achieving more than a sixfold reduction in energy consumption.
\end{abstract}
\section{Introduction}
\label{sec:intro}
Cochlear implants (CI) are a revolutionary medical device that allow the restoration of hearing in people with moderate to severe hearing loss \cite{kral-2021}. While effective in quiet settings, CI users struggle to understand speech in noisy environments due to the absence of temporal fine structure and spectral smearing caused by the implant \cite{fu_noise_2005}.

Traditional single channel speech enhancement algorithms such as time-frequency masking \cite{chiea_optimal_2021}, spectral subtraction \cite{yang_spectral_2005} and subspace algorithms \cite{loizou_subspace_2005} have been developed to address CI limitations. While they offer good performance in the presence of stationary noise, their performance degrades when exposed to non-stationary noise types \cite{chen_evaluation_2015}.

More recently, deep neural networks (DNN) have proven their capacity for enhancing speech masked by non-stationary noises \cite{borjigin_deep_2024, mamun_speech_2024, gajecki_deep_2023}. Despite promising results, DNNs require a large amount of energy to function making them non-ideal for low-power processors such as those found in a CI.

In contrast to DNNs, spiking neural networks (SNN) are a type artificial neural network (ANN) more closely inspired by the biological brain. SNNs process information in a sparse, binary and event-based manner making them significantly less power demanding than their DNN equivalents. In fact, it has been shown that SNNs can achieve orders of magnitude less energy consumption than their DNN equivalent when programmed on specialized neuromorphic hardware \cite{davies_advancing_2021}. SNNs have already shown great promise in the field of speech enhancement \cite{riahi_single_2023, hao_when_2024, sun_dpsnn_2024, arnaud_yarga_end--end_2025}, but have yet to be applied directly to CIs. 

In this paper, we propose Spiking DeepACE\footnote{Source code available at https://github.com/NECOTIS/spiking-deep-ace}, an SNN inspired by Deep ACE \cite{gajecki_deep_2023} that integrates the speech denoising in the CI processing pipeline. This paper is organized as follows, Section \ref{sec:spiking-deep-ace} presents the relevant background information and the proposed Spiking Deep ACE algorithm. Section \ref{sec:experimental-setup} presents the experimental setup used to train and evaluate Spiking Deep ACE and Deep ACE for comparison. The results are then presented in Section \ref{sec:results-and-discussion}, followed by a conclusion in Section \ref{sec:conclusion}.
\section{Spiking Deep ACE} \label{sec:spiking-deep-ace}
\begin{figure*}[t]
  \centering
  \includegraphics[width=\textwidth]{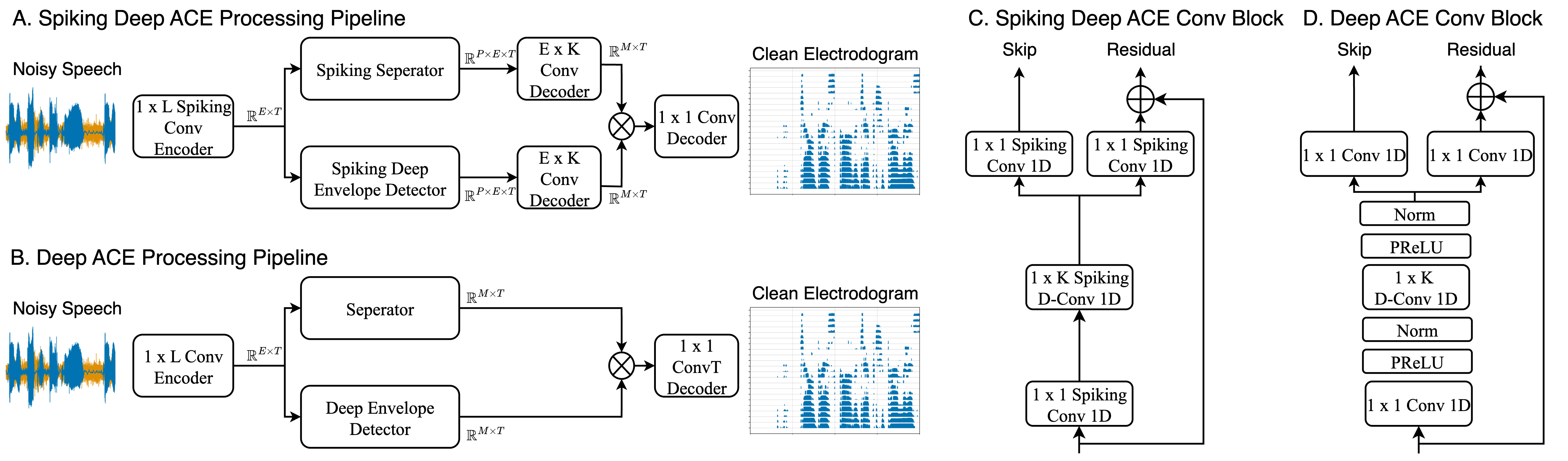}
  \caption{A) Spiking Deep ACE processing pipeline. B) Deep ACE processing pipeline. C) Convolutional blocks used in Spiking Deep ACE. D) Convolutional blocks used in Deep ACE.}
  \label{fig:architecture}
\end{figure*}

\subsection{Deep ACE}
\begin{figure*}[t]
  \centering
  \includegraphics[width=\textwidth]{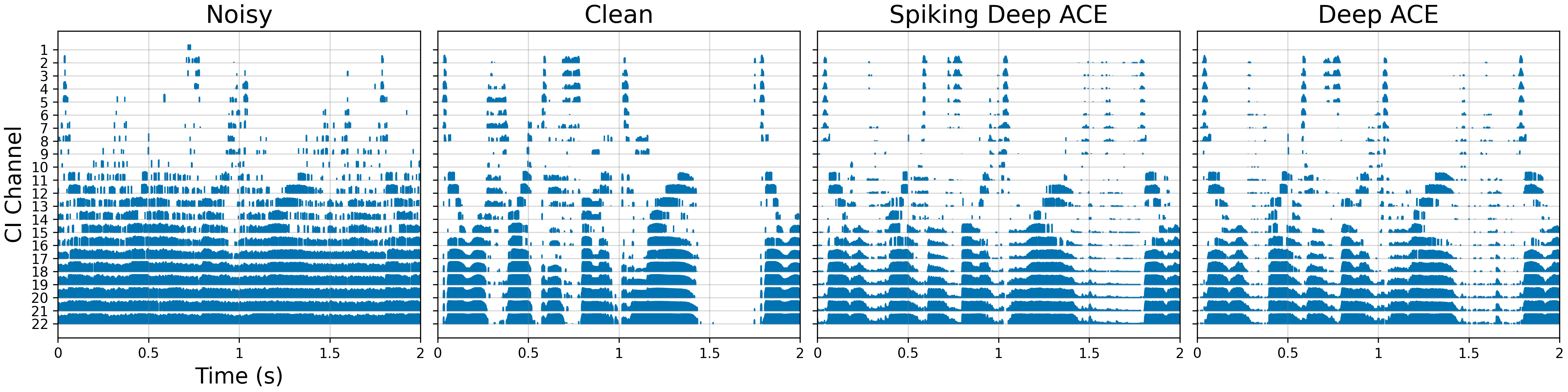}
  \caption{Example of predicted electrodograms by each of the models as well as the clean and noisy electrodograms for reference.}
  \label{fig:electrodograms}
\end{figure*}

The Deep ACE architecture was created in order to integrate speech denoising into the advanced combination encoder (ACE) sound coding strategy \cite{nogueira_psychoacoustic_2005} (see Figure\hyperref[fig:architecture]{~\ref*{fig:architecture}~(B)}). It is built upon the Conv-TasNet architecture for speaker separation \cite{luo_conv-tasnet_2019}. The main additions were an antirectifier unit in the encoder, a deep envelope detector (DED) in the skip path and a change in the output dimensions of the decoder. This allowed Deep ACE to efficiently denoise speech and predict the correct CI stimulation patterns while maintaining the \SI{2}{\ms} algorithmic latency of the standard ACE coding strategy \cite{gajecki_deep_2023}. These CI stimulation patterns can be represented with electrodograms in which the intensity of stimulation through time is represented for each CI channel (see Figure \ref{fig:electrodograms}).

\subsection{Spiking Neurons}
SNNs are composed of neuron models that more closely mimic the biological neuron than traditional ANN neurons. One of most popular model is the Leaky Integrate and Fire (LIF) model due to its simplicity \cite{izhikevich_which_2004}. In addition to its synaptic weights, the LIF model contains two other learnable parameters. The \emph{membrane leak constant} determines the rate at which the membrane potential decays back to its resting state and the \emph{threshold} determines the level of the membrane potential which causes an output spike.

Recently, a variation of the LIF was proposed by \cite{arnaud_yarga_end--end_2025} known as the ParaLIF. The ParaLIF seperates the membrane potential dynamics from the spike emission by removing the reset mecanism. This allows for parallel processing over time, providing orders of magnitude gains in training time compared to previous approaches \cite{arnaud_yarga_accelerating_2025}. In this study, we use the ParaLIF-Threshold variant of the ParaLIF \cite{arnaud_yarga_accelerating_2025}.

\subsection{Spiking Deep ACE}
Figure \ref{fig:architecture} presents the processing pipeline for Spiking Deep ACE and highlights the three major differences between the proposed Spiking Deep ACE and Deep ACE. Firstly, the complexity of the convolutional blocks was greatly reduced (see Figure\hyperref[fig:architecture]{~\ref*{fig:architecture}~(C)} and\hyperref[fig:architecture]{~\ref*{fig:architecture}~(D)}). In fact, preliminary experiments showed that the normalization layers did not provide any benefit to Spiking Deep ACE. Moreover, the spiking functions in ParaLIF neurons already act as a non-linearity removing the need for activation functions such as the PReLU used in Deep ACE.

Finally, Spiking Deep ACE's seperator and DED outputs are upsampled by a factor of $P$. This increases the number of features usable by the convolutional decoders to create the mask and the time-frequency representation allowing for superior denoising capabilities in comparison to no upsampling.

\section{Experimental Setup} \label{sec:experimental-setup}

\subsection{Datasets}
\label{sec-dataset}
The dataset used to train and test the models in this study is a publicly available dataset presented in \cite{valentini-dataset, valentini-botinhao_speech_2016}. The training set contains approximately 10 hours of noisy speech from 28 speakers (14 males and 14 females) at SNR levels of \SI{0}{\dB}, \SI{5}{\dB}, \SI{10}{\dB} and \SI{15}{\dB} as well as the corresponding clean speech. We randomly mixed the speakers from the training set and split them to have 80\% and 20\% of the speakers of each sex in the training and validation sets respectively. All audio data was then resampled to \SI{16}{\kHz} and split into two-second segments.

To test the models, we used synthesized unmodulated random Gaussian noise (ICRA static) and synthesized 6-speaker babble noise (ICRA babble) \cite{Dreschler2001ICRA}, following \cite{gajecki_deep_2023}. These noise signals were mixed with the clean testing set speech from \cite{valentini-dataset} at SNR levels of \SI{-5}{\dB}, \SI{0}{\dB}, \SI{5}{\dB} and \SI{10}{\dB} using the code from \cite{loizou-speech-enhancement}.

\subsection{Model Training}
Table \ref{table:hyperparameters} summarizes the hyperparameters used to train Deep ACE and Spiking Deep ACE. To ensure a valid comparison between Deep ACE and Spiking Deep ACE, Deep ACE was trained on the dataset described in section \ref{sec-dataset} with the same hyperparameters as in \cite{gajecki_deep_2023}. Given the additional neuronal parameters in the ParaLIF model and the greater number of neurons in the separator and DED outputs of Spiking Deep ACE than Deep ACE, a hyperparameter optimization was run in order to find the values $B$, $Sc$, $H$, $X$, $R$ that result in the best performance on the validation set for Spiking Deep ACE, while maintaining a parameter count less than or equal to that of Deep ACE.

\begin{table}[h]
    \centering
        \begin{tabular}{lll}
         Hyperparameter & Deep ACE & Spiking Deep ACE \\
         \hline
         N & 64 & 128 \\
         L & 32 & 32 \\
         B & 64 & 64 \\
         Sc & 32 & 64 \\
         H & 128 & 64 \\
         P & 3 & 3 \\
         X & 3 & 6 \\
         R & 8 & 4 \\
         \#Parameters & 552k & 437k \\
         \hline
        \end{tabular}
        \caption{Hyperparameters used in Deep ACE and Spiking Deep ACE.}
    \label{table:hyperparameters}
\end{table}

In terms of loss functions, Deep ACE uses a combination of a mean square error (MSE) between the clean electrodogram and the decoder output and a binary cross-entropy (BCE) between the ideal mask and the seperator output \cite{gajecki_deep_2023}. For Spiking Deep ACE, we only used the MSE loss as this was found to provide better results. Both models used the Adam optimizer \cite{kingma_adam_2017} with a learning rate of $10^{-3}$ and a learning rate scheduler to decrease the learning rate if the performance did not improve by a factor of $10^{-4}$ for 5 epochs.

\subsection{Performance Evaluation}
The first metric used to evaluate the model is the signal-to-noise ratio improvement (SNRi)  \cite{gajecki_deep_2023}. SNRi is computed in the electrodogram domain and measures the difference in SNR between the noisy electrodogram and the denoised electrodogram \cite{gajecki_deep_2023}.

The second metric used is the vocoded short-time objective intelligibility (VSTOI) \cite{watkins_predicting_2022}. VSTOI is based on the STOI metric \cite{taal_short-time_2010} and is used to estimate the intelligibility of the speech predicted by the models. In order to compute VSTOI, the predicted electrodogram was first re-synthesized as in \cite{gajecki_deep_2023}. The STOI was then computed using the synthesized speech and the clean unprocessed speech as the reference signal.
\begin{figure*}[!t]
  \centering
  \begin{minipage}[b]{0.48\textwidth}
    \centering
    \includegraphics[width=\linewidth]{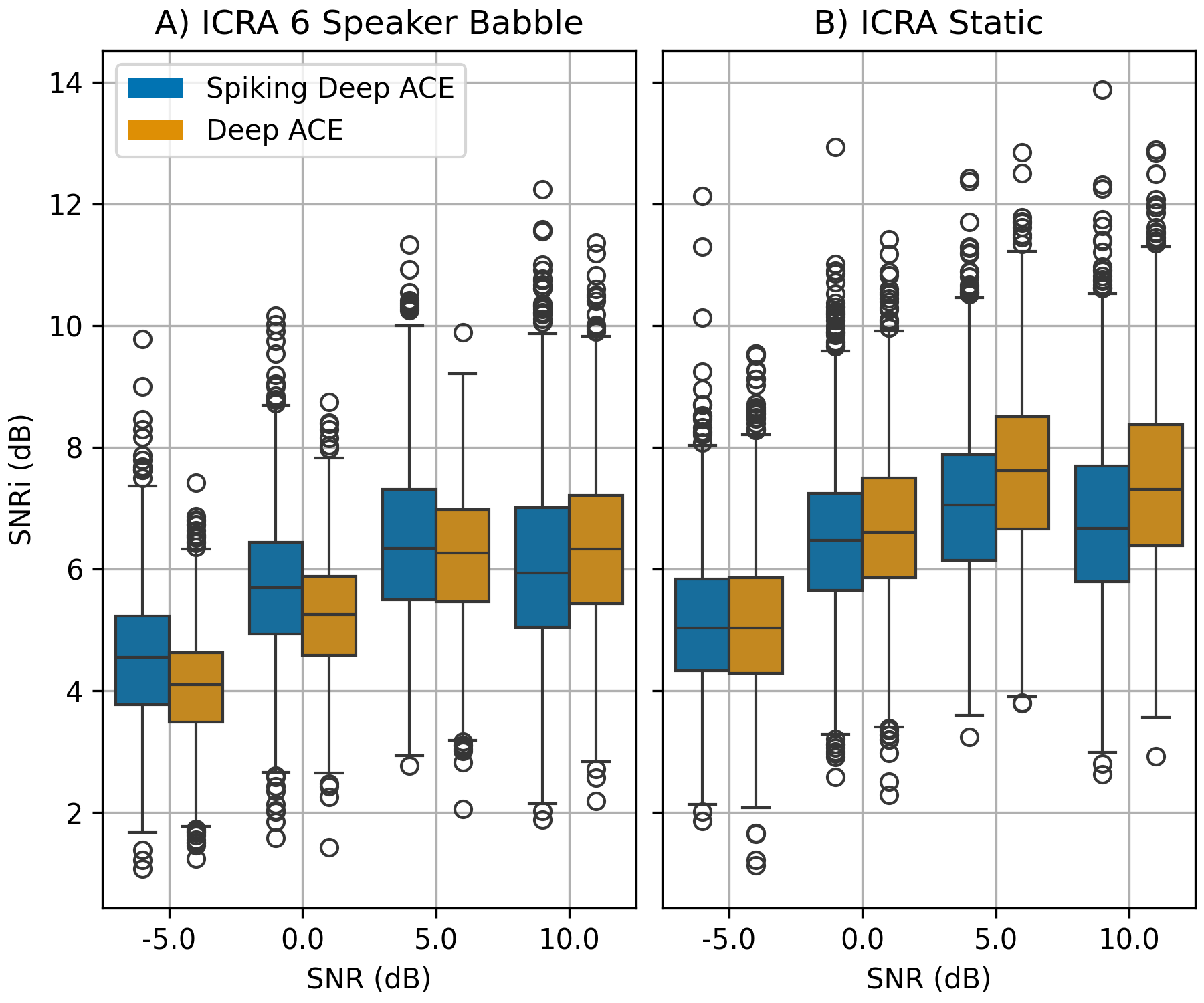}
    \centerline{(a)}
  \end{minipage}
  \hfill
  \begin{minipage}[b]{0.48\textwidth}
    \centering
    \includegraphics[width=\linewidth]{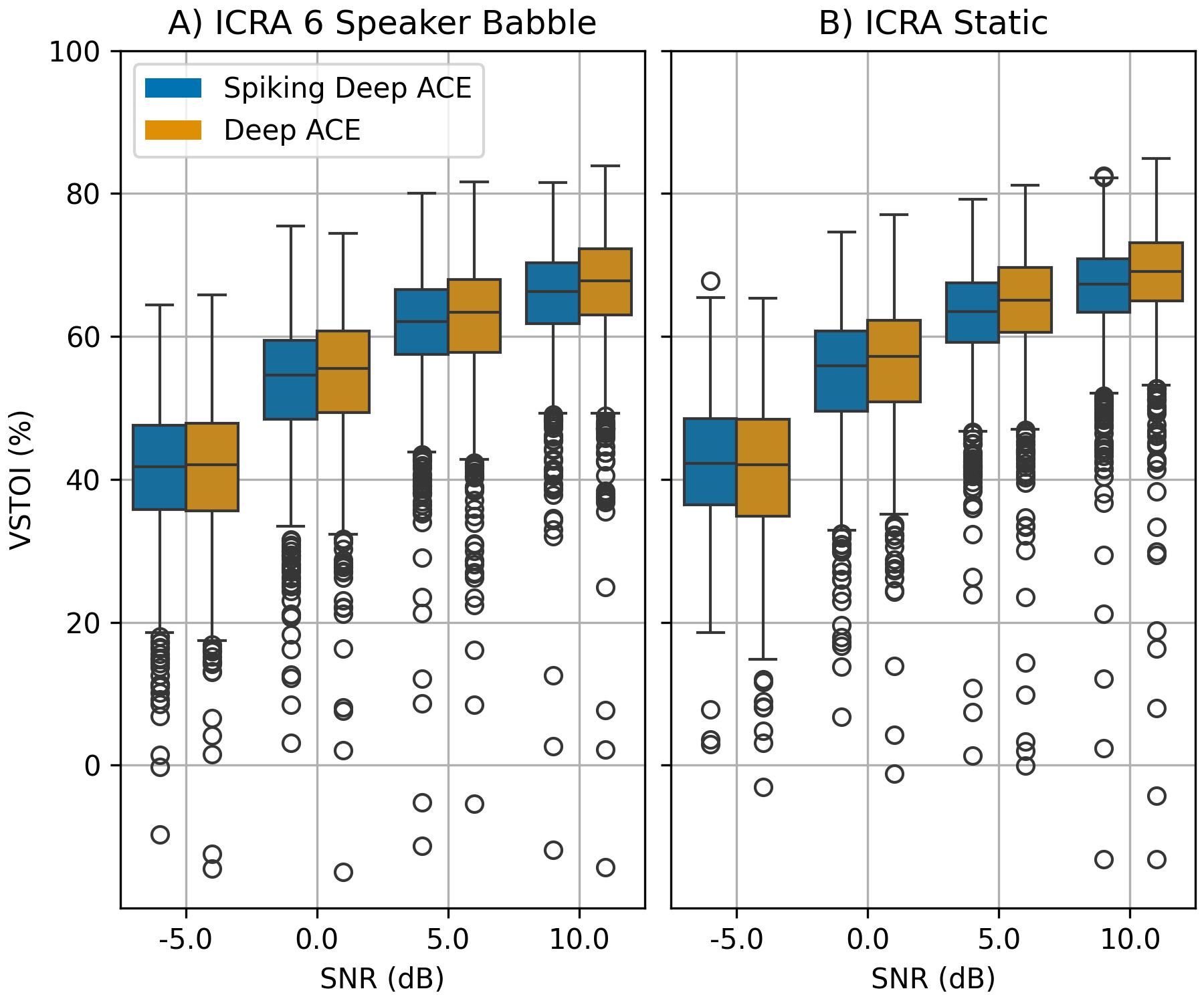}
    \centerline{(b)}
  \end{minipage}
  \caption{a) Box plots showing the VSTOI scores for both models on the ICRA babble and ICRA static test sets. b) Box plots showing the SNRi scores for both models on the same test sets as in a). In both a) and b), the center bar in each box represents the median score. The top and bottom edges of the box represent the third and first quartile respectively. The box whiskers extend to 1.5 times the inter-quartile range (IQR). Circles indicate results outside of 1.5 times the IQR.}
  \label{fig:results}
\end{figure*}

\subsection{Energy Evaluation}
The method to estimate the energy consumption of both models is based on the one used by \cite{riahi_single_2023}. The number of operations required for a convolutional layer is given by equation \ref{eq:ops},
\begin{equation} \label{eq:ops}
O = \frac{C_{in}}{G} \times K_{W} \times K_{H} \times C_{out} \times H_{out} \times T_{out}
\end{equation}
where $O$ is the number of operations, $C_{in}$ is the number of input channels, $G$ is the number of groups, $K_{W}$ is the kernel width, $K_H$ is the kernel height, $C_{out}$ is the number of output channels, $H_{out}$ is the height of the output and $T_{out}$ is the number of time steps in the output.
Since ANNs perform a multiply-and-accumulate (MAC) operation at every time step, the energy required for the synaptic operations in a layer is given by equation \ref{eq:energy_ann},
\begin{equation} \label{eq:energy_ann}
E_{ANN} = O \times C_{MAC} 
\end{equation}
where $E_{ANN}$ is the energy for the ANN layer and $C_{MAC}$ costs of a MAC operation on a CMOS \SI{45}{\nm} technology \cite{horowitz_11_2014}.

On the other hand, ParaLIF neurons only perform a floating point addition when there is an input spike at a given time step, but still need to update their internal state at every time step. This update costs an order of magnitude more than a floating point addition \cite{timcheck_intel_2023}. Therefore, the energy required for an SNN layer is given by equation \ref{eq:energy_snn_full},
\begin{equation} \label{eq:energy_snn_full}
E_{SNN} = \frac{\overline{S}}{T_{out}} \times O \times C_{ADD} + T_{out} \times N \times 10 \times C_{ADD}
\end{equation}
where $E_{SNN}$ represents the energy for the SNN layer, $\overline{S}$ is the mean input spike rate for the layer, $C_{ADD}$ is the cost of a floating point addition operation on a CMOS \SI{45}{\nm} technology \cite{horowitz_11_2014} and $N$ is the number of neurons in the layer.

The total energy of the network is computed by summing equation \ref{eq:energy_ann} over all layers that process floating point values, and equation  \ref{eq:energy_snn_full} over all layers that process spikes.

\section{Results and Discussion}
\label{sec:results-and-discussion}
\subsection{Performance Results}
Figure \ref{fig:results} presents the SNRi and VSTOI scores obtained for both Spiking Deep ACE and Deep ACE on the ICRA static and ICRA babble noise types across various SNR levels. The results show that Spiking Deep ACE is able to achieve competitive performance with Deep ACE accoss all noise types and SNR levels. In fact, Spiking Deep ACE outperforms Deep ACE in the lower SNRs of the ICRA static test set. In the other cases, Spiking Deep ACE's mean SNRi stays well within \SI{0.5}{\dB} of Deep ACE's SNRi.

Similarly, Spiking Deep ACE's VSTOI scores stays competitive with those of Deep ACE in all SNRs and noise types. In fact, Spiking Deep ACE's VSTOI stays within 2\% of Deep ACE's VSTOI in all of the SNR levels.

\subsection{Energy Results}
Table \ref{table:results} shows the considerable energy gains that Spiking Deep ACE achieves over Deep ACE. Notably, Spiking Deep ACE maintains competitive performance with Deep ACE while consuming more than 6 times less energy.
\begin{figure}[t]
  \centering
  \includegraphics[width=8.5cm]{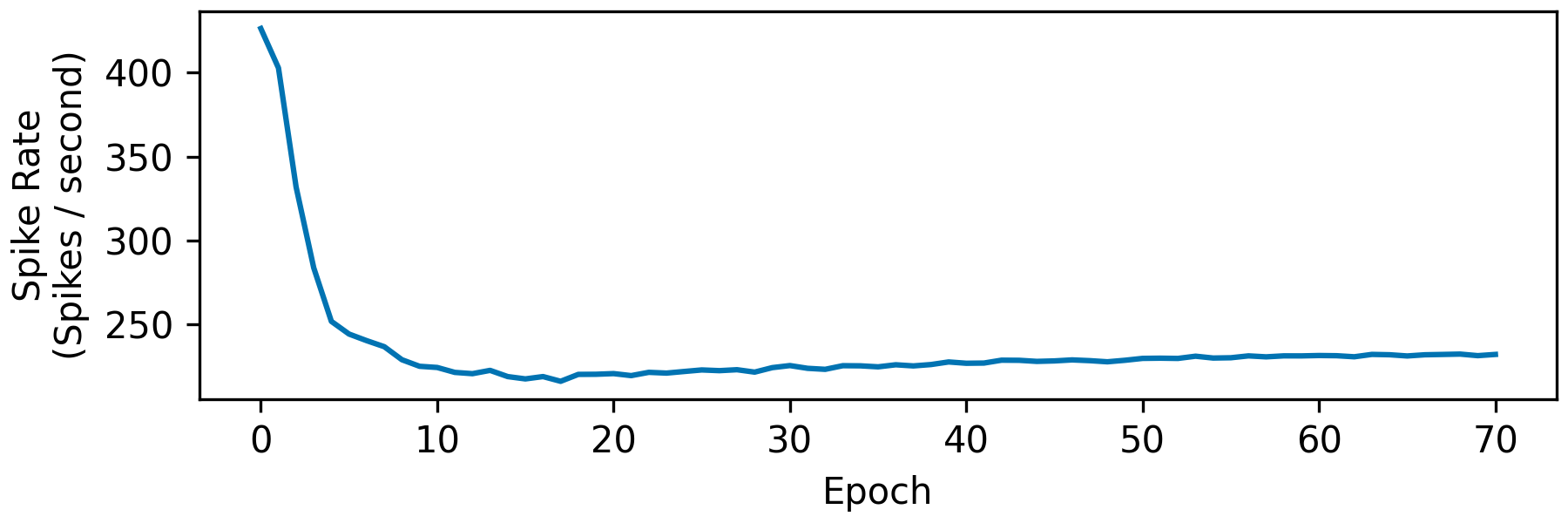}
  \caption{Evolution of the mean spike rate per neuron during training. Since the sampling rate of the network is \SI{1}{\kHz}, a spike rate of 250 spikes / second implies one spike every 4 time steps.}
  \label{fig:sparsity}
\end{figure}

\begin{table}[h]
    \centering
        \begin{tabular}{ p{3cm} p{1cm} p{1cm} p{1cm} }
         Model & VSTOI (\%) & SNRi (dB) & Energy ($\mu J / s$) \\
         \hline
         Spiking Deep ACE & 56 & 6.0 & \textbf{372} \\
         Deep ACE & \textbf{57} & \textbf{6.1} & 2461 \\
         \hline
        \end{tabular}
        \caption{Results and energy comparison between Deep ACE and Spiking Deep ACE on the ICRA babble and ICRA static test sets combined.}
    \label{table:results}
\end{table}

We note that equation \ref{eq:energy_snn_full} does not include the energy costs for data transfers to and from memory. These transactions account for the large majority of energy in ANNs and SNNs \cite{han_learning_2015}. However, in addition to having less parameters than Deep ACE, Spiking Deep ACE is intrinsically sparse (see Figure \ref{fig:sparsity}). Thus, Spiking Deep ACE requires less memory transfers than Deep ACE allowing for additional energy gains.
\section{Conclusion}
\label{sec:conclusion}
In this paper, we have proposed an SNN capable of end-to-end speech denoising and CI coding. The proposed Spiking Deep ACE model was compared to the Deep ACE model in terms of VSTOI, SNRi and energy consumption. It was shown that our approach achieved competitive results in comparison to Deep ACE while consuming considerably less energy than Deep ACE, highlighting its potential for low-power CI processors.

\bibliographystyle{IEEEbib}
\bibliography{strings,refs}

@book{kral-2021,
    author = "Kral, A. and Aplin, F. and Maier H.",
    title = "Prostheses for the Brain: Introduction to Neuroprosthetics",
    publisher = "Academic Press",
    year = 2021,
    address = "London, United Kingdom",
    pages = 414,

}

@book{loizou-speech-enhancement,
    author = "Loizou, P. C.",
    title = "SPEECH ENHANCEMENT: Theory and Practice",
    publisher = "CRC Press",
    year = 2013,
    address = "Boca Raton",
    pages = "632",
}

@misc{valentini-dataset,
    title = {Noisy speech database for training speech enhancement algorithms and {TTS} models},
    author = {Valentini-Botinhao, C.},
    year = 2017,
    publisher = {University of Edinburgh. School of Informatics. Centre for Speech Technology Research (CSTR)},
    doi = {https://doi.org/10.7488/ds/2117}
}

@inproceedings{valentini-botinhao_speech_2016,
	title = {Speech {Enhancement} for a {Noise}-{Robust} {Text}-to-{Speech} {Synthesis} {System} {Using} {Deep} {Recurrent} {Neural} {Networks}},
	url = {https://www.isca-archive.org/interspeech_2016/valentinibotinhao16_interspeech.html},
	doi = {10.21437/Interspeech.2016-159},
	language = {en},
	urldate = {2025-09-04},
	booktitle = {Interspeech 2016},
	publisher = {ISCA},
	author = {Valentini-Botinhao, Cassia and Wang, Xin and Takaki, Shinji and Yamagishi, Junichi},
	month = sep,
	year = {2016},
	pages = {352--356},
}

@article{Dreschler2001ICRA,
  author    = {Dreschler, W. A. and Verschuure, H. and Ludvigsen, C. and Westermann, S.},
  title     = {{ICRA} noises: artificial noise signals with speech-like spectral and temporal properties for hearing instrument assessment. International Collegium for Rehabilitative Audiology},
  journal   = {Audiology},
  year      = {2001},
}

@article{nogueira_psychoacoustic_2005,
	title = {A {Psychoacoustic} ``{NofM}"-{Type} {Speech} {Coding} {Strategy} for {Cochlear} {Implants}},
	volume = {2005},
	issn = {1687-6180},
	url = {https://asp-eurasipjournals.springeropen.com/articles/10.1155/ASP.2005.3044},
	doi = {10.1155/ASP.2005.3044},
	language = {en},
	number = {18},
	urldate = {2024-09-20},
	journal = {EURASIP Journal on Advances in Signal Processing},
	author = {Nogueira, W. and Büchner, A. and Lenarz, T. and Edler, B.},
	month = dec,
	year = {2005},
	pages = {101672},
}

@article{gajecki_deep_2023,
	title = {A {Deep} {Denoising} {Sound} {Coding} {Strategy} for {Cochlear} {Implants}},
	volume = {70},
	copyright = {https://creativecommons.org/licenses/by/4.0/legalcode},
	issn = {0018-9294, 1558-2531},
	url = {https://ieeexplore.ieee.org/document/10083222/},
	doi = {10.1109/TBME.2023.3262677},
	language = {en},
	number = {9},
	urldate = {2024-09-20},
	journal = {IEEE Transactions on Biomedical Engineering},
	author = {Gajecki, T. and Zhang, Y. and Nogueira, W.},
	month = sep,
	year = {2023},
	pages = {2700--2709},
}

@article{borjigin_deep_2024,
	title = {Deep learning restores speech intelligibility in multi-talker interference for cochlear implant users},
	volume = {14},
	issn = {2045-2322},
	url = {https://www.nature.com/articles/s41598-024-63675-8},
	doi = {10.1038/s41598-024-63675-8},
	language = {en},
	number = {1},
	urldate = {2024-09-20},
	journal = {Scientific Reports},
	author = {Borjigin, A. and Kokkinakis, K. and Bharadwaj, H. M. and Stohl, J. S.},
	month = jun,
	year = {2024},
	pages = {13241},
}

@article{mamun_speech_2024,
	title = {Speech {Enhancement} for {Cochlear} {Implant} {Recipients} {Using} {Deep} {Complex} {Convolution} {Transformer} {With} {Frequency} {Transformation}},
	volume = {32},
	copyright = {https://ieeexplore.ieee.org/Xplorehelp/downloads/license-information/IEEE.html},
	issn = {2329-9290, 2329-9304},
	url = {https://ieeexplore.ieee.org/document/10443501/},
	doi = {10.1109/TASLP.2024.3366760},
	language = {en},
	urldate = {2024-09-25},
	journal = {IEEE/ACM Transactions on Audio, Speech, and Language Processing},
	author = {Mamun, N. and Hansen, J. H. L.},
	year = {2024},
	pages = {2616--2629},
}

@inproceedings{riahi_single_2023,
	address = {Regina, SK, Canada},
	title = {Single {Channel} {Speech} {Enhancement} {Using} {U}-{Net} {Spiking} {Neural} {Networks}},
	copyright = {https://doi.org/10.15223/policy-029},
	isbn = {979-8-3503-2397-9},
	url = {https://ieeexplore.ieee.org/document/10288830/},
	doi = {10.1109/CCECE58730.2023.10288830},
	language = {en},
	urldate = {2024-09-25},
	booktitle = {2023 {IEEE} {Canadian} {Conference} on {Electrical} and {Computer} {Engineering} ({CCECE})},
	publisher = {IEEE},
	author = {Riahi, A. and Plourde, É.},
	month = sep,
	year = {2023},
	pages = {111--116},
}

@article{chiea_optimal_2021,
	title = {An {Optimal} {Envelope}-{Based} {Noise} {Reduction} {Method} for {Cochlear} {Implants}: {An} {Upper} {Bound} {Performance} {Investigation}},
	volume = {29},
	copyright = {https://ieeexplore.ieee.org/Xplorehelp/downloads/license-information/IEEE.html},
	issn = {2329-9290, 2329-9304},
	shorttitle = {An {Optimal} {Envelope}-{Based} {Noise} {Reduction} {Method} for {Cochlear} {Implants}},
	url = {https://ieeexplore.ieee.org/document/9420255/},
	doi = {10.1109/TASLP.2021.3076363},
	language = {en},
	urldate = {2024-09-26},
	journal = {IEEE/ACM Transactions on Audio, Speech, and Language Processing},
	author = {Chiea, R. A. and Costa, M. H. and Cordioli, J. A.},
	year = {2021},
	pages = {1729--1739},
}

@inproceedings{hao_when_2024,
	address = {Singapore, Singapore},
	title = {When {Audio} {Denoising} {Meets} {Spiking} {Neural} {Network}},
	copyright = {https://doi.org/10.15223/policy-029},
	isbn = {979-8-3503-5409-6},
	url = {https://ieeexplore.ieee.org/document/10605482/},
	doi = {10.1109/CAI59869.2024.00275},
	language = {en},
	urldate = {2024-10-06},
	booktitle = {2024 {IEEE} {Conference} on {Artificial} {Intelligence} ({CAI})},
	publisher = {IEEE},
	author = {Hao, X. and Ma, C. and Yang, Q. and Tan, K. C. and Wu, J.},
	month = jun,
	year = {2024},
	pages = {1524--1527},
}

@article{chen_evaluation_2015,
	title = {Evaluation of {Noise} {Reduction} {Methods} for {Sentence} {Recognition} by {Mandarin}-{Speaking} {Cochlear} {Implant} {Listeners}},
	volume = {36},
	issn = {0196-0202},
	url = {https://journals.lww.com/00003446-201501000-00007},
	doi = {10.1097/AUD.0000000000000074},
	language = {en},
	number = {1},
	urldate = {2024-10-08},
	journal = {Ear \& Hearing},
	author = {Chen, F. and Hu, Y. and Yuan, M.},
	month = jan,
	year = {2015},
	pages = {61--71},
}

@article{izhikevich_which_2004,
	title = {Which {Model} to {Use} for {Cortical} {Spiking} {Neurons}?},
	volume = {15},
	copyright = {https://ieeexplore.ieee.org/Xplorehelp/downloads/license-information/IEEE.html},
	issn = {1045-9227},
	url = {http://ieeexplore.ieee.org/document/1333071/},
	doi = {10.1109/TNN.2004.832719},
	language = {en},
	number = {5},
	urldate = {2024-10-11},
	journal = {IEEE Transactions on Neural Networks},
	author = {Izhikevich, E. M.},
	month = sep,
	year = {2004},
	pages = {1063--1070},
}

@article{davies_advancing_2021,
	title = {Advancing {Neuromorphic} {Computing} {With} {Loihi}: {A} {Survey} of {Results} and {Outlook}},
	volume = {109},
	copyright = {https://creativecommons.org/licenses/by/4.0/legalcode},
	issn = {0018-9219, 1558-2256},
	shorttitle = {Advancing {Neuromorphic} {Computing} {With} {Loihi}},
	url = {https://ieeexplore.ieee.org/document/9395703/},
	doi = {10.1109/JPROC.2021.3067593},
	language = {en},
	number = {5},
	urldate = {2024-10-15},
	journal = {Proceedings of the IEEE},
	author = {Davies, M. and Wild, A. and Orchard, G. and Sandamirskaya, Y. and Guerra, G. A. F. and Joshi, P. and Plank, P. and Risbud, S. R.},
	month = may,
	year = {2021},
	pages = {911--934},
}

@article{luo_conv-tasnet_2019,
	title = {Conv-{TasNet}: {Surpassing} {Ideal} {Time}–{Frequency} {Magnitude} {Masking} for {Speech} {Separation}},
	volume = {27},
	copyright = {https://ieeexplore.ieee.org/Xplorehelp/downloads/license-information/OAPA.html},
	issn = {2329-9290, 2329-9304},
	shorttitle = {Conv-{TasNet}},
	url = {https://ieeexplore.ieee.org/document/8707065/},
	doi = {10.1109/TASLP.2019.2915167},
	language = {en},
	number = {8},
	urldate = {2024-11-29},
	journal = {IEEE/ACM Transactions on Audio, Speech, and Language Processing},
	author = {Luo, Y. and Mesgarani, N.},
	month = aug,
	year = {2019},
	pages = {1256--1266},
}

@article{yang_spectral_2005,
	title = {Spectral subtraction-based speech enhancement for cochlear implant patients in background noise},
	volume = {117},
	issn = {0001-4966, 1520-8524},
	url = {https://pubs.aip.org/jasa/article/117/3/1001/544207/Spectral-subtraction-based-speech-enhancement-for},
	doi = {10.1121/1.1852873},
	language = {en},
	number = {3},
	urldate = {2024-11-27},
	journal = {The Journal of the Acoustical Society of America},
	author = {Yang, L. and Fu, Q.},
	month = mar,
	year = {2005},
	pages = {1001--1004},
}

@article{loizou_subspace_2005,
	title = {Subspace algorithms for noise reduction in cochlear implants},
	volume = {118},
	issn = {0001-4966, 1520-8524},
	url = {https://pubs.aip.org/jasa/article/118/5/2791/897003/Subspace-algorithms-for-noise-reduction-in},
	doi = {10.1121/1.2065847},
	language = {en},
	number = {5},
	urldate = {2024-11-27},
	journal = {The Journal of the Acoustical Society of America},
	author = {Loizou, P. C. and Lobo, A. and Hu, Y.},
	month = nov,
	year = {2005},
	pages = {2791--2793},
}

@article{arnaud_yarga_end--end_2025,
	title = {End-to-end neuromorphic speech enhancement with {PDM} microphones$^{\textrm{*}}$},
	volume = {5},
	issn = {2634-4386},
	url = {https://iopscience.iop.org/article/10.1088/2634-4386/adf2d4},
	doi = {10.1088/2634-4386/adf2d4},
	language = {en},
	number = {3},
	urldate = {2025-08-01},
	journal = {Neuromorphic Computing and Engineering},
	author = {A. Yarga, S. Y. and Wood, Sean U. N.},
	month = sep,
	year = {2025},
	pages = {034009},
}

@article{sun_dpsnn_2024,
	title = {{DPSNN}: spiking neural network for low-latency streaming speech enhancement},
	volume = {4},
	issn = {2634-4386},
	shorttitle = {{DPSNN}},
	url = {https://iopscience.iop.org/article/10.1088/2634-4386/ad93f9},
	doi = {10.1088/2634-4386/ad93f9},
	language = {en},
	number = {4},
	urldate = {2025-08-01},
	journal = {Neuromorphic Computing and Engineering},
	author = {Sun, T. and Bohté, S.},
	month = dec,
	year = {2024},
	pages = {044008},
}

@article{arnaud_yarga_accelerating_2025,
	title = {Accelerating spiking neural networks with parallelizable leaky integrate-and-fire neurons$^{\textrm{*}}$},
	volume = {5},
	issn = {2634-4386},
	url = {https://iopscience.iop.org/article/10.1088/2634-4386/adb7fe},
	doi = {10.1088/2634-4386/adb7fe},
	language = {en},
	number = {1},
	urldate = {2025-09-04},
	journal = {Neuromorphic Computing and Engineering},
	author = {A. Yarga, S. Y. and Wood, Sean U. N.},
	month = mar,
	year = {2025},
	pages = {014012},
}

@inproceedings{taal_short-time_2010,
	address = {Dallas, TX, USA},
	title = {A short-time objective intelligibility measure for time-frequency weighted noisy speech},
	copyright = {https://doi.org/10.15223/policy-029},
	isbn = {978-1-4244-4295-9},
	url = {https://ieeexplore.ieee.org/document/5495701/},
	doi = {10.1109/ICASSP.2010.5495701},
	language = {en},
	urldate = {2025-09-10},
	booktitle = {2010 {IEEE} {International} {Conference} on {Acoustics}, {Speech} and {Signal} {Processing}},
	publisher = {IEEE},
	author = {Taal, C. H. and Hendriks, R. C. and Heusdens, R. and Jensen, J.},
	month = mar,
	year = {2010},
	pages = {4214--4217},
}

@misc{kingma_adam_2017,
	title = {Adam: {A} {Method} for {Stochastic} {Optimization}},
	shorttitle = {Adam},
	url = {http://arxiv.org/abs/1412.6980},
	doi = {10.48550/arXiv.1412.6980},
	language = {en},
	urldate = {2025-09-14},
	publisher = {arXiv},
	author = {Kingma, D. P. and Ba, J.},
	month = jan,
	year = {2017},
	note = {arXiv:1412.6980 [cs]},
}

@article{watkins_predicting_2022,
	title = {Predicting {Speech} {Intelligibility} for {Individual} {Cochlear} {Implant} {Recipients}},
	volume = {75},
	issn = {0745-7472},
	url = {https://journals.lww.com/10.1097/01.HJ.0000831136.08368.17},
	doi = {10.1097/01.HJ.0000831136.08368.17},
	language = {en},
	number = {5},
	urldate = {2025-09-14},
	journal = {The Hearing Journal},
	author = {Watkins, G.},
	month = may,
	year = {2022},
	pages = {20,21,24,25,26},
}

@article{timcheck_intel_2023,
	title = {The {Intel} neuromorphic {DNS} challenge},
	volume = {3},
	issn = {2634-4386},
	url = {https://iopscience.iop.org/article/10.1088/2634-4386/ace737},
	doi = {10.1088/2634-4386/ace737},
	language = {en},
	number = {3},
	urldate = {2025-09-14},
	journal = {Neuromorphic Computing and Engineering},
	author = {Timcheck, J. and Shrestha, S. B. and B. D. Rubin, D. and Kupryjanow, A. and Orchard, G. and Pindor, L. and Shea, T. and Davies, M.},
	month = sep,
	year = {2023},
	pages = {034005},
}

@article{fu_noise_2005,
	title = {Noise {Susceptibility} of {Cochlear} {Implant} {Users}: {The} {Role} of {Spectral} {Resolution} and {Smearing}},
	volume = {6},
	copyright = {http://www.springer.com/tdm},
	issn = {1525-3961, 1438-7573},
	shorttitle = {Noise {Susceptibility} of {Cochlear} {Implant} {Users}},
	url = {http://link.springer.com/10.1007/s10162-004-5024-3},
	doi = {10.1007/s10162-004-5024-3},
	language = {en},
	number = {1},
	urldate = {2025-09-15},
	journal = {Journal of the Association for Research in Otolaryngology},
	author = {Fu, Q. and Nogaki, G.},
	month = mar,
	year = {2005},
	pages = {19--27},
}

@inproceedings{horowitz_11_2014,
	address = {San Francisco, CA, USA},
	title = {1.1 {Computing}'s energy problem (and what we can do about it)},
	isbn = {978-1-4799-0920-9 978-1-4799-0918-6},
	url = {http://ieeexplore.ieee.org/document/6757323/},
	doi = {10.1109/ISSCC.2014.6757323},
	language = {en},
	urldate = {2025-09-16},
	booktitle = {2014 {IEEE} {International} {Solid}-{State} {Circuits} {Conference} {Digest} of {Technical} {Papers} ({ISSCC})},
	publisher = {IEEE},
	author = {Horowitz, M.},
	month = feb,
	year = {2014},
	pages = {10--14},
}

@article{han_learning_2015,
	title = {Learning both {Weights} and {Connections} for {Efficient} {Neural} {Network}},
	language = {en},
    journal = {Advances in Neural Information Processing Systems},
    year = {2015},
    volume = {28},
	author = {Han, S. and Pool, J. and Tran, J. and Dally, W.},
}

\end{document}